\documentclass[twocolumn,letter]{jpsj3}
\usepackage{color}
\usepackage{braket}
\usepackage{mathrsfs}
\usepackage{graphicx}
\usepackage{graphics}
\usepackage{amssymb, latexsym}
\usepackage{amsmath}
\usepackage{bm}
\usepackage{color}
\usepackage{ulem}
\usepackage{fancybox}
\usepackage{xspace}
\usepackage{amsfonts}
\usepackage{multirow}
\usepackage[section]{placeins}
\usepackage{dcolumn}
\usepackage{amsthm}
\usepackage{setspace}
\usepackage{dsfont}

\def\runauthor{Y. Akagi, H. Katsura, and T. Koma}

\title{A New Numerical Method for $\mathbb{Z}_2$ Topological Insulators with Strong Disorder}

\author{Yutaka Akagi$^1$\thanks{E-mail address: akagi@cams.phys.s.u-tokyo.ac.jp}, Hosho Katsura$^1$, and Tohru Koma$^2$
}
\inst{$^1$Department of Physics, Graduate School of Science, The University of Tokyo, Hongo, Tokyo 113-0033, Japan\\
$^2$Department of Physics, Gakushuin University, Mejiro, Toshima-ku, Tokyo 171-8588, Japan
}
\abst{We propose a new method to numerically compute the $\mathbb{Z}_2$ indices
for disordered topological insulators in Kitaev's periodic table.
All of the $\mathbb{Z}_2$ indices are known to be derived from the index formulae
which are expressed in terms of a pair of projections 
introduced by Avron, Seiler, and Simon. For a given pair of projections, the corresponding index is
determined by the spectrum of the difference between the two projections.
This difference exhibits remarkable and useful properties, as it is compact and has a supersymmetric structure in the spectrum. These properties make it possible to numerically determine the indices of disordered topological insulators highly efficiently. 
The method is demonstrated for the Bernevig-Hughes-Zhang and Wilson-Dirac models whose topological phases are characterized by a $\mathbb{Z}_2$ index in two and three dimensions, respectively. 
}

\kword{Topological insulators, Time reversal symmetry, $\mathbb{Z}_2$ topological invariant, disordered systems}

\begin{document}
\maketitle
Since the discovery of $\mathbb{Z}_2$ topological insulators by Kane and Mele \cite{Kane05}, many methods have been proposed to compute the $\mathbb{Z}_2$ index. 
In particular, Fu and Kane found that the calculation of the $\mathbb{Z}_2$ index can be considerably simplified in a system with inversion symmetry \cite{Fu07}. 
However, for disordered systems, a numerical determination of the $\mathbb{ Z}_2$ index is still very challenging.
Roughly speaking, three types of numerical methods have been proposed so far for this problem:

\noindent
$\bullet$ The first one was proposed 
by Kane and Mele themselves \cite{Kane05}, and later some 
modifications were introduced. 
Basically, for a given model, the $\mathbb{ Z}_2$ index is defined by a certain Pfaffian 
with twisted boundary conditions \cite{Niu85}. 
The methods were applied to class AII models \cite{Schnyder08,Kitaev09,Ryu10} 
in two and three dimensions with or without a certain 
inversion symmetry
\cite{Essin07,Guo10,Leung12}. 

\noindent
$\bullet$ The second one is based on a scattering matrix theory \cite{Fulga12}. The $\mathbb{ Z}_2$ indices which are 
defined by the scattering matrices were numerically computed for models in the classes, AII and DIII, 
in two and three dimensions \cite{Fulga12,Sbierski14}. 

\noindent
$\bullet$ The third one was proposed by Loring and Hastings \cite{Loring10}.  
The $\mathbb{ Z}_2$ indices are defined by Bott indices  which are introduced as an obstruction to 
approximating almost commuting matrices. 
For models in the class AII in two and three dimensions, the $\mathbb{ Z}_2$ indices were 
numerically computed \cite{Loring10,Loring15}.  
For systems with chiral symmetry, Loring and Schulz-Baldes \cite{Loring17} proposed a numerical method 
to obtain the values of Bott indices.

In this paper, we propose an alternative method 
to numerically calculate the $\mathbb{ Z}_2$ indices for disordered topological insulators in arbitrary dimensions. 
The method is based on the index formulae which were derived in Refs.~[15, 16]. 
and has the following two advantages: (i) There is no need to take an average of the $\mathbb{ Z}_2$ index over random variables in a model. 
(ii) The $\mathbb{ Z}_2$ index is determined by the discrete spectrum of a certain compact operator 
with a supersymmetric structure. 
The latter makes it possible to numerically determine the $\mathbb{Z}_2$ index highly efficiently.

Our method is demonstrated for Bernevig-Hughes-Zhang (BHZ) \cite{Bernevig06,Yamakage12,Yamakage13} and 
Wilson-Dirac \cite{Wilson74,Qi08,Kobayashi13} models whose topological phases are characterized by 
a $\mathbb{ Z}_2$ index of the class AII in two and three dimensions, respectively. 
In consequence, the method enables us to determine all of the values of the $\mathbb{ Z}_2$ indices 
of the strong and weak topological insulators, and the normal insulator phases in the phase diagrams. 
These values of the $\mathbb{ Z}_2$ indices 
completely coincide with the predictions 
in previous studies using a reliable transfer-matrix method \cite{Yamakage12,Yamakage13,Kobayashi13}.

\begin{figure}[t]
\begin{center}
\includegraphics[width=1.0\columnwidth]{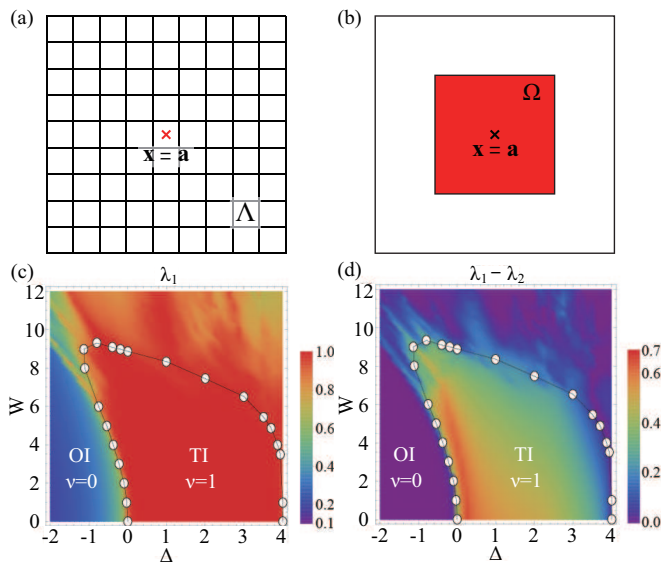}
\end{center}
\caption{(Color online).
(a) The location of ${\bm a}$ is chosen to be the center of the finite lattice $\Lambda$ of linear size $L$. 
(b) The subregion $\Omega$ is chosen so that its center is the same as ${\bm a}$ and it does not overlap with the boundary of $\Lambda$. 
(c) and (d), respectively, show $\lambda_1$ and \mbox{$\lambda_1 - \lambda_2$} in the BHZ model
as a function of the mass $\Delta$ and the disorder strength $W$. 
The obtained value $\nu$ of the $\mathbb{Z}_2$ index is indicated in both OI and TI phases. 
The values of the parameters used are \mbox{$E_{\rm F}=0$}, \mbox{$g_1=g_2=1$}, 
\mbox{$g_3=0$}, and the system size is \mbox{$L^2=1600$}. 
The curves of the phase boundaries with the dots are plotted by using the results in Ref.~[19]. 
}
\label{Fig:PD_BHZ}
\end{figure}

To begin with, we introduce two Dirac operators as \cite{Katsura16a,Katsura16b} 
\begin{equation}
\mathcal{D}_{\bm a}({\bm x}) := \frac{x_1 + i x_2 - (a_1 + i a_2)}{| x_1 + i x_2 - (a_1 + i a_2) |} 
\end{equation}
in two dimension (2D), and 
\begin{equation}
D_{\bm a}({\bm x}) := \frac{1}{|{\bm x} - {\bm a}|} ({\bm x} - {\bm a}) \cdot \mbox{\boldmath $\sigma$}
\end{equation}
in three dimensions (3D), where ${\bm x} =$ \mbox{$(x_1, \cdots, x_d)\in\mathbb{Z}^d$} is the position operator 
and ${\bm a} =$ \mbox{$(a_1, \cdots, a_d)\in \mathbb{R}^d \backslash \mathbb{Z}^d$} 
is a vector for $d=2,3$ [see Fig.~\ref{Fig:PD_BHZ}(a)]. 
The three-component vector $\mbox{\boldmath $\sigma$}$ is defined by 
$\mbox{\boldmath $\sigma$}=\mbox{$(\sigma_1, \sigma_2, \sigma_3)$}$ 
whose components are given by Pauli matrices $\sigma_i$, each of which acts on an auxiliary Hilbert space $\mathbb{C}^2$. 
The whole Hilbert space is given by the tensor product of the auxiliary $\mathbb{C}^2$ and the original Hilbert space for the Hamiltonian of tight-binding models which we will consider shortly.

Next, we define the $\mathbb{Z}_2$ index for an infinite-volume system which is a tight-binding model
on a square $\mathbb{Z}^2$ or cubic lattice $\mathbb{Z}^3$~\cite{comment}. 
Let $P_{\rm F}$ be the projection operator onto the states below the Fermi energy $E_{\rm F}$.
The difference of two projections is defined as \cite{Avron94b,Katsura16b} 
\begin{eqnarray}
A := \begin{cases}
    P_{\rm F} - \mathcal{D}_{\bm a}^* P_{\rm F} \mathcal{D}_{\bm a} & {\rm in \: \: 2D} \\
    P_{\rm F} - D_{\bm a} P_{\rm F}D_{\bm a} & {\rm in \: \: 3D},
  \end{cases}
\label{eq:A_op}
\end{eqnarray}
where $\mathcal{D}_{\bm a}^*$ is the adjoint of the Dirac operator $\mathcal{D}_{\bm a}$. 
Then, the $\mathbb{Z}_2$ index $\nu$ is defined as 
\begin{eqnarray}
  \nu = {\rm dim} \; {\rm ker}\; (A - 1) \; \; {\rm modulo} \; 2.
  \label{eq:Z2-index}
\end{eqnarray}
When the Fermi energy $E_{\rm F}$ lies in a spectral gap or a mobility gap, 
the $\mathbb{Z}_2$ index is known to be well defined \cite{Katsura16b}. 
In the following, we will consider such situations. 

Now we describe our numerical scheme for calculating the $\mathbb{Z}_2$ index. 
Let $\Lambda$ and $\Omega$ be two finite regions satisfying $\Omega\subset\Lambda\subset\mathbb{R}^d$ 
and ${\bm a}\in\Omega$ as in Fig.~\ref{Fig:PD_BHZ}. 
First, we approximate the Fermi projection $P_{\rm F}$ in the infinite volume by the corresponding Fermi projection in the finite region $\Lambda$. More precisely, the approximate one is given by   
\begin{eqnarray}
P_{\rm F}^{(\Lambda)}:= \sum_{E_n \le E_{\rm F}} \ket{n} \bra{n},
\end{eqnarray}
where $\ket{n}$ are eigenstates of the tight-binding Hamiltonian $\mathcal{H}$ on $\Lambda$ which we consider,  
and we have denoted by $E_n$ the corresponding eigenvalues. 
To avoid the presence of gapless edge/surface states, we impose periodic boundary conditions for the Hamiltonian in practical numerical calculations. 
For the operator $A$ in Eq.~(\ref{eq:A_op}), we replace the Fermi projection $P_{\rm F}$ by $P_{\rm F}^{(\Lambda)}$, 
and write $A^{(\Lambda)}$ for the approximate one.  
Further, we restrict the operator $A^{(\Lambda)}$ to the subregion $\Omega$ as 
\begin{equation}
\label{AOmegaLambda}
A_{\Omega}^{(\Lambda)} := \chi_\Omega A^{(\Lambda)} \chi_\Omega,
\end{equation}
where $\chi_\Omega$ is the characteristic function of the subregion $\Omega$.

Under the above gap assumption, the operator $A$ in Eq.~(\ref{eq:A_op}) is compact even in the infinite volume limit. 
Therefore, the spectrum of $A$ is discrete and has no accumulation point except for zero. 
This implies that an eigenstate of $A$ is localized if the corresponding eigenvalue is not equal to zero. 
Let $\lambda\ne 0$ be an eigenvalue of $A$. 
From these observations, it is clear that the eigenvalue $\lambda$ can be approximated by an eigenvalue $\lambda'$ of the approximate operator $A_\Omega^{(\Lambda)}$ if the subregion $\Omega$ is sufficiently large. 
In addition to this, the cutoff function $\chi_\Omega$ in Eq.~(\ref{AOmegaLambda}) enables us to remove 
the boundary effects due to the finite size of the region $\Lambda$.

As the first demonstration, we compute the $\mathbb{ Z}_2$ index of the BHZ model \cite{Bernevig06,Yamakage12,Yamakage13} 
with disorder on a square lattice $\mathbb{Z}^2$. The Hamiltonian is written as
\begin{eqnarray}
    \mathcal{H}_{\rm 2D} \! = \! \sum_{\bm x} \! \sum_{k=1,2} \! \bigl[ c^{\dagger}_{\bm x} t_k c_{{\bm x}
+{\bm e}_k} \! + \!{\rm h.c.}  \bigr]
                \! + \! \sum_{\bm x} c^{\dagger}_{\bm x} (\tau_0 \! \otimes \! \epsilon_{\bm x}) c_{\bm x},
\label{Eq:BHZ}
\end{eqnarray}
where $c_{\bm x}=$ \mbox{$[c_{{\bm x} 1 \uparrow}, c_{{\bm x} 1 \downarrow}, c_{{\bm x} 2 \uparrow}, c_{{\bm x} 2 \downarrow}]^T$}, 
and $c_{{\bm x} i \alpha}$ is the annihilation operator of an electron with spin $\alpha$ and orbital $i$ 
at site ${\bm x}$. 
The hopping matrices, $t_1$ and $t_2$, are given by $t_1= g_1\tau_0 \otimes s_3 -\frac{i}{2} g_2 \tau_0 \otimes s_2 + \frac{i}{2} g_3 \tau_2 \otimes s_3$, and $t_2= g_1\tau_0 \otimes s_3 -\frac{i}{2} g_2 \tau_3 \otimes s_1 - \frac{i}{2} g_3 \tau_2 \otimes s_0$,
where $\tau_k$ and $s_k$ \mbox{($k=1,2,3$)} are Pauli matrices for the orbital and the spin, respectively. 
Here, $g_1, g_2$ and $g_3$ are real parameters. 
The on-site potential $\epsilon_{\bm x}$ is given by 
$\epsilon_{\bm x}=$ \mbox{${\rm diag}[\Delta - 4 g_1 + W_{\bm x}^{+}, - \Delta + 4 g_1 + W_{\bm x}^{-}]$}, 
where $W_{\bm x}^{+}$ and $W_{\bm x}^{-}$ are a random potential whose distribution is uniform in the interval \mbox{$[-W/2, W/2]$} with a positive parameter $W$.
${\bm e}_1$ and ${\bm e}_2$ are the unit vectors in the $x$ and $y$ directions, respectively.
As is well known \cite{Bernevig06,Yamakage12,Yamakage13}, this Hamiltonian (\ref{Eq:BHZ}) belongs to the symmetry class AII. 
We set the Fermi energy \mbox{$E_{\rm F}=0$} which is located at the center of the energy gap.

In the following, we write $\lambda_i$ for the {\it i}-th eigenvalue of the operator $A_{\Omega}^{(\Lambda)}$
of Eq.~(\ref{AOmegaLambda}) in descending order including multiplicity~\cite{comment2}. 
We note that 
the spectrum of $A_{\Omega}^{(\Lambda)}$ is included in the interval $[-1,1]$. 

Before showing our numerical results, 
we abbreviate the topological and the ordinary insulator phases as TI and OI, respectively, 
in the phase diagram \cite{Yamakage12,Yamakage13}, and write $\nu$ for the value of the $\mathbb{ Z}_2$ index.
Figure \ref{Fig:PD_BHZ}(c) and (d) show, respectively, $\lambda_1$ and \mbox{$\lambda_1 - \lambda_2$} 
as a function of the mass $\Delta$ and the strength  $W$ of disorder.
In the region TI, our numerical results are satisfactory 
because \mbox{$\lambda_1 \simeq 1$} and \mbox{$\lambda_1 - \lambda_2 \neq 0$}. 
Actually, these imply $\nu=1$, i.e., the phase is topological as we expected.
In the region OI, $\lambda_1$ is significantly different from 1, and thus $\nu=0$.
These results show that the $\mathbb{ Z}_2$ index enables us to distinguish TI from OI. 
In OI phase, 
one notices $\lambda_1\simeq\lambda_2\lesssim 0.2$ as seen in Fig.~\ref{Fig:PD_BHZ}(d). 
This degeneracy is nothing but a consequence of the two symmetries \cite{Katsura16a,Katsura16b}, 
the time-reversal symmetry of the Hamiltonian and 
the supersymmetric structure of the operator $A$. This kind of degeneracy is very useful 
to determine the $\mathbb{ Z}_2$ index.  
In the region 
$W \lesssim 6$, our results are in good agreement with 
the previous results \cite{Yamakage13} which were obtained by using a transfer-matrix method. 
In the region with a sufficiently large $W$, Anderson localization is expected to occur. 
For the intermediate critical region between these two regions, our method 
is not under control
because the existence of a significantly nonvanishing spectral or mobility gap cannot be expected. 
In fact, our numerical results in this region show large fluctuations in both $\lambda_1$ and $\lambda_1-\lambda_2$. 

The second example is the Wilson-Dirac model \cite{Kobayashi13} with disorder on the cubic lattice $\mathbb{Z}^3$.
The Hamiltonian is written as 
\begin{eqnarray}
    \mathcal{H}_{\rm 3D} = \mathcal{H}_{\sf 0} + \mathcal{H}_{\sf hop} + \mathcal{H}_{\sf dis},
\label{Eq:WD}
\end{eqnarray}
where
\begin{eqnarray}
    \mathcal{H}_{\sf 0} \! \! \! \! \! &=& \! \! \! \! \! \sum_{\bm x} \! \! \sum_{k=1,2,3} \! \bigl[\frac{i t}{2} c^{\dagger}_{{\bm x}+{\bm e}_k} {\alpha}_k c_{\bm x} 
                         \! - \! \frac{m_2}{2} c^{\dagger}_{{\bm x}+{\bm e}_k} \beta c_{\bm x} \! + \! {\rm h.c.} \bigr] \notag\\ 
                         &+& \! \! \! \! \! (m+3m_2) \sum_{\bm x} c^{\dagger}_{\bm x} \beta c_{\bm x}, \\
    \mathcal{H}_{\sf hop} \! \! \! \! \! &=& \! \! \! \! \! \sum_{\bm x} \sum_{k=1,2,3} \bigl[ t_0 c^{\dagger}_{{\bm x}+{\bm e}_k} c_{\bm x} + {\rm h.c.} \bigr], \\
    \mathcal{H}_{\sf dis} \! \! \! \! \! &=& \! \! \! \! \! \sum_{\bm x} v_{\bm x} c^{\dagger}_{\bm x} c_{\bm x}.
\end{eqnarray}
Here,
$c_{\bm x}=$ \mbox{$[c_{{\bm x} 1 \uparrow}, c_{{\bm x} 1 \downarrow}, c_{{\bm x} 2 \uparrow}, c_{{\bm x} 2 \downarrow}]^T$},
the vector ${\bm e}_k$ is the unit vector in $k=x,y,z$ direction,
and $\alpha_k =$ \mbox{$\tau_1 \otimes s_k$}, $\beta =$ \mbox{$\tau_3 \otimes s_0$}; 
$v_{\bm x}$ is the on-site random potential whose 
distribution is uniform in the interval \mbox{$[-W/2, W/2]$} with a positive parameter $W$. 
This Hamiltonian (\ref{Eq:WD}) 
belongs to the symmetry class AII for \mbox{$W \neq 0$} or \mbox{$t_0 \neq 0$}. 
In the following, we will treat the case with \mbox{$t_0 \neq 0$}.

\begin{figure}[t]
\begin{center}
\includegraphics[width=1.0\columnwidth]{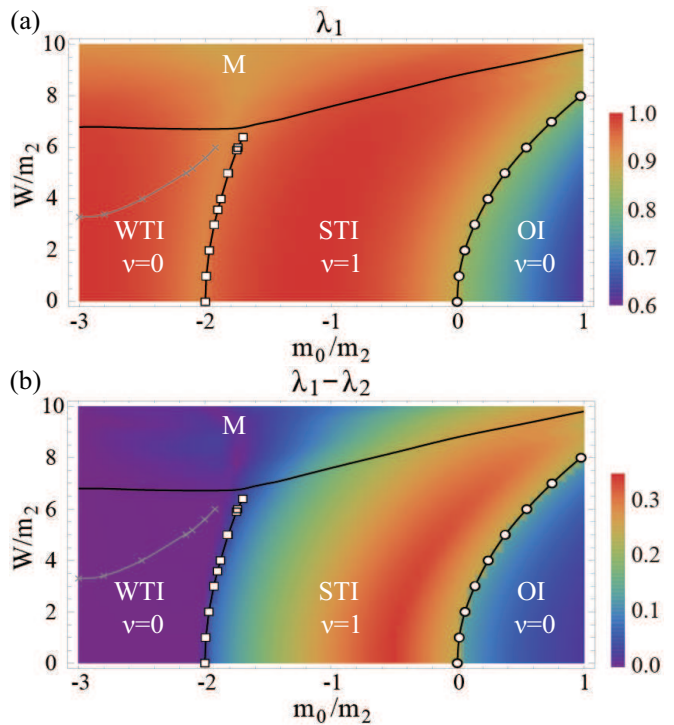}
\end{center}
\caption{(Color online).
(a) and (b), respectively, show $\lambda_1$ and \mbox{$\lambda_1 - \lambda_2$} in the Wilson-Dirac-type model as a function of the mass parameter $m_0/m_2$ and the disorder strength $W/m_2$. 
The numerical values $\nu$ of the $\mathbb{ Z}_2$ index are indicated in the phases WTI, STI and OI. 
The parameters used are 
\mbox{$E_{\rm F}=0$}, \mbox{$t=2$}, \mbox{$t_0=0.01$}, and the system size is \mbox{$L^3=1728$}.
The curves of the phase boundaries with the dots are plotted by using the results in Ref.~[22].
}
\label{Fig:PD_Wilson-Dirac}
\end{figure}

\begin{figure}[t]
\begin{center}
\includegraphics[width=0.8\columnwidth]{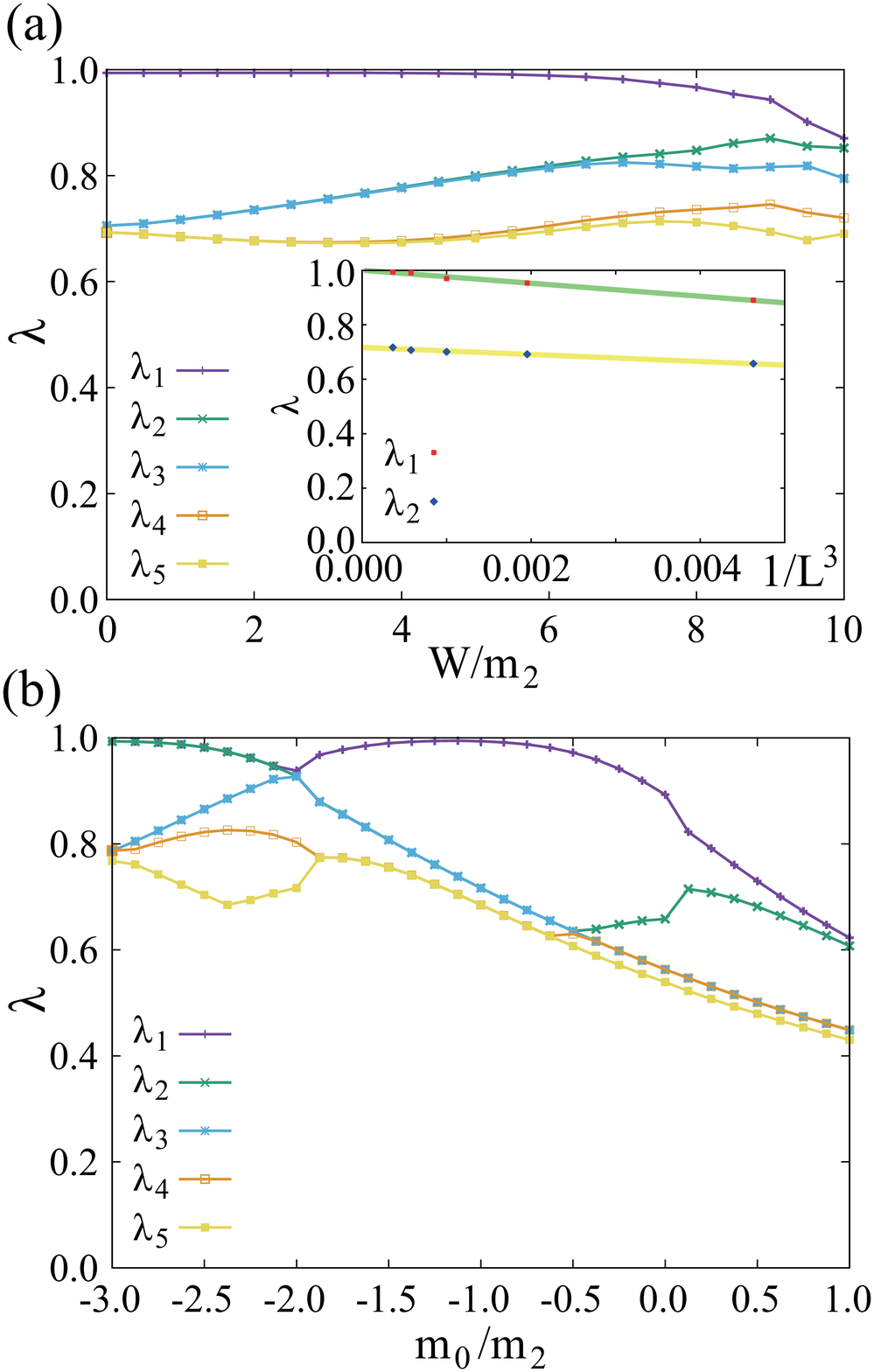}
\end{center}
\caption{(Color online).
(a) $W/m_2$ and (b) $m_0/m_2$ dependences of the eigenvalues $\lambda_i$, $i=1,2,\ldots,5$, 
of the operator $A_{\Omega}^{(\Lambda)}$ for the system size \mbox{$L^3=2744$}.
(a) and (b) correspond to 
cross-sections of Fig.~\ref{Fig:PD_Wilson-Dirac} for \mbox{$m_0/m_2=-1.0$} and \mbox{$W/m_2=1.0$}, respectively.
The inset in (a) shows 
$\lambda_1$ and $\lambda_2$ 
as a function of the inverse volume of the system ($1/L^3$) for fixed parameters, 
\mbox{$m_0/m_2=-1.0$} and \mbox{$W/m_2=1.0$}. 
The green (yellow) line is a fit to 
the numerical data 
of $\lambda_1$ ($\lambda_2$). 
}
\label{Fig:cross-section}
\end{figure}

Similarly, we abbreviate the weak, strong topological, the ordinary insulator and the diffusive metal phases 
as WTI, STI, OI, and M, respectively \cite{Kobayashi13}, and write $\nu$ for the value of the $\mathbb{ Z}_2$ index~\cite{comment3}. 
Figure~\ref{Fig:PD_Wilson-Dirac} shows $\lambda_1$ and \mbox{$\lambda_1 - \lambda_2$} 
as a function of the mass parameter $m_0/m_2$ and the strength $W/m_2$ of disorder. 
In the region OI, $\nu=0$ because 
\mbox{$\lambda_1 \lesssim 0.8$}. 
In the region STI, \mbox{$\lambda_1 \simeq 1$} and \mbox{$\lambda_1 - \lambda_2 \neq 0$}, 
and hence $\nu=1$. 
In the region WTI, 
\mbox{$\lambda_1 = \lambda_2 \simeq 1$} but $\lambda_3$ is significantly different from 1
[see Fig.~\ref{Fig:cross-section}(b)],
and hence $\nu=0$ because the multiplicity of the eigenvalue \mbox{$\lambda  \simeq  1$} is two.
As for the region M, 
we cannot expect our method to be effective
because the spectral or mobility gap 
is expected to vanish if the metallic character of the spectrum is true. 
To summarize, as seen in Fig.~\ref{Fig:PD_Wilson-Dirac}, our numerical results for the $\mathbb{ Z}_2$ index 
are in good agreement with the predictions of Ref.~[22]. 
In particular, the phase boundaries between WTI and STI, and between STI and OI are considerably sharp. 
Moreover, although our method is expected to be useless
in the metallic phase, 
there do not appear fluctuations like those in the two-dimensional case.

Although the $\mathbb{Z}_2$ index vanishes in both the OI and WTI phases, there is a definite difference between them as follows: OI phase yields no eigenvalue \mbox{$\lambda \simeq 1$} 
while WTI phase yields the eigenvalue \mbox{$\lambda \simeq 1$} which is doubly degenerate.
If the eigenvalue \mbox{$\lambda \simeq 1$} is related to surface states, 
one can expect, from our numerical results, that the multiplicity of \mbox{$\lambda \simeq 1$}  
coincides with the number of Dirac cones which appear on the surface of the system. 
In fact, it was numerically observed 
in a generalized Kane-Mele model on a diamond lattice
that WTI (STI) phase exhibits two surface Dirac cones (odd number of Dirac cones)~\cite{Fu07b}.

Figure~\ref{Fig:cross-section} shows $W/m_2$ and $m_0/m_2$ dependences of the eigenvalues $\lambda_i$, $i=1,2,\ldots,5$, 
of the operator $A_{\Omega}^{(\Lambda)}$.
According to Ref.~[22], 
STI emerges for \mbox{$W/m_2 \lesssim 7$} and for \mbox{$m_0/m_2 = -1.0$}. 
One can see that $\lambda_1$ is significantly different from $\lambda_2$, and 
that $\lambda_2$ and $\lambda_3$ are degenerate, and similarly, $\lambda_4$ and $\lambda_5$ are degenerate.  
As mentioned in the case of two dimensions, 
this even degeneracy is a consequence of the two symmetries \cite{Katsura16a,Katsura16b}, 
the time-reversal symmetry of the Hamiltonian and the supersymmetric structure of the operator $A$.
For the region \mbox{$W/m_2 \gtrsim 7$} in Fig.~\ref{Fig:cross-section}(a), 
the diffusive metallic phase appears.
In this region, one can see that the above double degeneracy in the spectrum of $A_{\Omega}^{(\Lambda)}$ is lifted  
due to the vanishing of the spectral or mobility gap in the metallic phase.

The inset of Fig.~\ref{Fig:cross-section}(a) shows system-size dependences
of $\lambda_1$ 
and $\lambda_2$ for fixed parameters \mbox{$m_0/m_2=-1.0$} and \mbox{$W/m_2=1.0$}. 
As the system size increases, 
$\lambda_1$ and $\lambda_2$ converge to $1$ and about $0.7$, respectively.
Thus in the infinite-volume limit,
we can clearly conclude that the $\mathbb{ Z}_2$ index $\nu$ is unity.
One can perform a similar analysis and extract the eigenvalues of $A$ in the infinite-volume limit for other values of the parameters.

As seen in Fig.~\ref{Fig:cross-section}(b), the double degeneracy of \mbox{$\lambda \simeq 1$} appears 
in the region of WTI where the corresponding parameter satisfies \mbox{$m_0/m_2 \lesssim -2$}.
Clearly, one can see that $\lambda_3$ is separated from \mbox{$\lambda_1 \simeq \lambda_2 \simeq 1$},
and hence the $\mathbb{ Z}_2$ index $\nu$ is equal to zero.
Thus, if the first and the second eigenvalues, $\lambda_1$ and $\lambda_2$, are degenerate, 
information about the third eigenvalue $\lambda_3$ is useful to determine the $\mathbb{ Z}_2$ index. 
\smallskip

Summary: We have presented our new method for numerically calculating the $\mathbb{ Z}_2$ indices 
of topological insulators with strong disorder.  
Our method has the following two advantages:   
(i) There is no need to take an average of the $\mathbb{ Z}_2$ index over random variables 
in a model. (ii) The $\mathbb{ Z}_2$ index is determined by the discrete spectrum of a certain compact operator 
with a supersymmetric structure. 
These properties make it possible to numerically determine the $\mathbb{Z}_2$ index highly efficiently. 
In order to check the effectiveness of our method, we have demonstrated that 
all of the numerical values of the $\mathbb{ Z}_2$ indices 
completely coincide with the predictions 
in previous studies using a reliable transfer-matrix method \cite{Yamakage12,Yamakage13,Kobayashi13}  
for the two-dimensional Bernevig-Hughes-Zhang and the three-dimensional Wilson-Dirac models.
Thus, the strong topological insulator phases can be characterized by the $\mathbb{ Z}_2$ index 
in the index formulae \cite{Katsura16a,Katsura16b}, and can be 
clearly distinguished from other phases. 
We believe that the good agreement between the two different approaches is one of the steps 
toward the understanding of the nature of $\mathbb{Z}_2$ topological insulators 
although we cannot definitely compare our method with other approaches mentioned 
at the beginning of the present paper.
Finally, we remark that the generalization of our method to models in other symmetry classes in arbitrary dimensions is straightforward.

\acknowledgements
{
The authors acknowledge helpful discussions with Ken-Ichiro Imura, Takahiro Misawa, Tomi Ohtsuki and Synge Todo.
This work was supported by 
JSPS KAKENHI Grant Nos. JP15K17719, JP16H00985, JP17K14352.
}


\begin{thebibliography}{101}

\bibitem{Kane05}
C. L. Kane and E. J. Mele,
\textit{${{ Z}}_2$ Topological Order and the Quantum Spin ${{\rm H}}$all Effect}, 
Phys. Rev. Lett. \textbf{95}, 146802 (2005).

\bibitem{Fu07}
L. Fu and C. L. Kane,
\textit{Topological insulators with inversion symmetry}, 
Phys. Rev. B \textbf{76}, 045302 (2007).

\bibitem{Niu85}
Q. Niu, D. J. Thouless, and Y.-S. Wu,
\textit{Quantized ${{\rm H}}$all conductance as a topological invariant}, 
Phys. Rev. B \textbf{31}, 3372 (1985).

\bibitem{Schnyder08}
A. P. Schnyder, S. Ryu, A. Furusaki, and A. W. W. Ludwig,
\textit{Classification of topological insulators and superconductors in three spatial dimensions}, 
Phys. Rev. B \textbf{78}, 195125 (2008).

\bibitem{Kitaev09}
A. Kitaev,
\textit{Periodic table for topological insulators and superconductors}, 
AIP Conference Proceedings \textbf{1134}, 22 (2009).

\bibitem{Ryu10}
S. Ryu, A. P. Schnyder, A. Furusaki, and A. W. W. Ludwig,
\textit{Topological insulators and superconductors: tenfold way and dimensional hierarchy}, 
New J. Phys. \textbf{12}, 065010 (2010).

\bibitem{Essin07}
A. M. Essin and J. E. Moore,
\textit{Topological insulators beyond the ${{\rm B}}$rillouin zone via ${{\rm C}}$hern parity}, 
Phys. Rev. B \textbf{76}, 165307 (2007).

\bibitem{Guo10}
H.-M. Guo,
\textit{Topological invariant in three-dimensional band insulators with disorder}, 
Phys. Rev. B \textbf{82}, 115122 (2010).

\bibitem{Leung12}
B. Leung and E. Prodan,
\textit{Effect of strong disorder in a three-dimensional topological insulator: 
Phase diagram and maps of the ${{ Z}}_2$ invariant}, 
Phys. Rev. B \textbf{85}, 205136 (2012).

\bibitem{Fulga12}
I. C. Fulga, F. Hassler, and A. R. Akhmerov,
\textit{Scattering theory of topological insulators and superconductors}, 
Phys. Rev. B \textbf{85}, 165409 (2012).

\bibitem{Sbierski14}
B. Sbierski and P. W. Brouwer,
\textit{${{ Z}}_2$ phase diagram of three-dimensional disordered topological insulators via a scattering matrix approach}, 
Phys. Rev. B \textbf{89}, 155311 (2014).

\bibitem{Loring10}
T. A. Loring and M. B. Hastings,
\textit{Disordered topological insulators via ${C}^*$-algebras}, 
EPL (Europhysics Lett.) \textbf{92}, 67004 (2010).

\bibitem{Loring15}
T. A. Loring,
\textit{K-theory and pseudospectra for topological insulators}, 
Ann. Phys. \textbf{356}, 383 (2015).

\bibitem{Loring17}
T. A. Loring and H. Schulz-Baldes,
\textit{Finite volume calculation of {K-theory} invariants}, 
Preprint arXiv:1701.07455 (2017).

\bibitem{Katsura16a}
H. Katsura and T. Koma,
\textit{The ${{ Z}}_2$ index of disordered topological insulators with time reversal symmetry}, 
J. Math. Phys. \textbf{57}, 021903 (2016).

\bibitem{Katsura16b}
H. Katsura and T. Koma,
\textit{The Noncommutative Index Theorem and the Periodic Table for Disordered Topological Insulators and Superconductors}, 
Preprint arXiv:1611.01928 (2016).

\bibitem{Bernevig06}
B. A. Bernevig, T. L. Hughes, and S.-C. Zhang,
\textit{Quantum Spin ${{\rm H}}$all Effect and Topological Phase Transition in ${{\rm H}}$g${{\rm T}}$e Quantum Wells}, 
Science \textbf{314}, 1757 (2006).

\bibitem{Yamakage12}
A. Yamakage, K. Nomura, K.-I. Imura, and Y. Kuramoto,
\textit{${{ Z}}_2$ Topological Anderson Insulator}, 
J. Phys.: Conf. Ser. \textbf{400}, 042070 (2012).

\bibitem{Yamakage13}
A. Yamakage, K. Nomura, K.-I. Imura, and Y. Kuramoto,
\textit{Criticality of the metal-topological insulator transition driven by disorder}, 
Phys. Rev. B \textbf{87}, 205141 (2013).

\bibitem{Wilson74}
K. G. Wilson,
\textit{Confinement of quarks}, 
Phys. Rev. D \textbf{10}, 2445 (1974).

\bibitem{Qi08}
X.-L. Qi, T. L. Hughes, and S.-C. Zhang,
\textit{Topological field theory of time-reversal invariant insulators}, 
Phys. Rev. B \textbf{78}, 195424 (2008).

\bibitem{Kobayashi13}
K. Kobayashi, T. Ohtsuki, and K.-I. Imura,
\textit{Disordered Weak and Strong Topological Insulators}, 
Phys. Rev. Lett. \textbf{110}, 236803 (2013).

\bibitem{comment}
We note that a tight-binding model on an arbitrary lattice in $d$ dimension can be mapped onto the model on $\mathbb{Z}^d$ with suitably chosen hopping integrals.

\bibitem{Avron94b}
J. Avron, R. Seiler, and B. Simon,
\textit{The Index of a Pair of Projections}, 
J. Func. Anal. \textbf{120}, 220 (1994).

\bibitem{comment2}
Here, $A^{(\Lambda)}$ is expressed in terms of $P_{\rm F}$ and $\tilde{\mathcal{D}}_a({\bm x}) :=$ \mbox{$\mathcal{D}_a({\bm x}) \otimes {\mathbb I}_{4}$} 
as $A^{(\Lambda)}=P_{\rm F} - {\tilde {\cal D}}^*_a P_{\rm F} {\tilde {\cal D}}_a$, where ${\mathbb I}_{4}$ is the 4-dimensional identity matrix 
and $\mathcal{D}_a$ should be thought of as a matrix of dimension $|\Lambda|$.

\bibitem{comment3}
In this case, the operator $A^{(\Lambda)}$ is written as $A^{(\Lambda)} = \tilde{P}_{\rm F} - \tilde{D}_a \tilde{P}_{\rm F} \tilde{D}_a $, 
where $\tilde{P}_{\rm F} := P_{\rm F}  \otimes {\mathbb I}_{2}$, $\tilde{D}_a({\bm x}) := \sum_{i=1}^3$ \mbox{$D_a^i({\bm x})  \otimes {\mathbb I}_{4}  \otimes \sigma_i$}, and 
$D_a^i({\bm x}) :=  (x_i - a_i)/|{\bm x} - {\bm a}|$. 
Here, ${\mathbb I}_{2}$ and ${\mathbb I}_{4}$ are the 2- and 4-dimensional identity matrices, respectively. 
Each operator $D_a^i$ should be thought of as a matrix of dimension $|\Lambda|$.

\bibitem{Fu07b}
L. Fu, C. L. Kane, and E. J. Mele,
\textit{Topological Insulators in Three Dimensions}, 
Phys. Rev. Lett. \textbf{98}, 106803 (2007).



\end{thebibliography}
\end{document}